\documentclass[12pt,a4paper]{article}
\pdfoutput=1
\usepackage[utf8]{inputenc}
\usepackage[T1]{fontenc}
\usepackage[margin=1.9cm]{geometry}
\usepackage[pdfstartview={FitH},bookmarks=false,linktoc=page,colorlinks=true,linkcolor=blue,citecolor=blue,urlcolor=blue]{hyperref}
\usepackage[numbered]{bookmark}
\usepackage{mathtools}
\usepackage{amssymb}
\usepackage{graphicx}
\usepackage[font=small,labelfont=bf]{caption}
\usepackage{authblk}
\usepackage{cite}
\usepackage{dsfont}
\usepackage{xcolor}
\usepackage[titles]{tocloft}
\usepackage{float}
\usepackage{subcaption}
\usepackage{color}
\usepackage{tikz}
\usepackage{ifthen}
\usepackage[warn]{textcomp}
\numberwithin{equation}{section}
\allowdisplaybreaks
\usepackage{multirow}

\usetikzlibrary{patterns}

\usepackage{comment}

\usepackage{caption}
\usepackage{subcaption}

\begin{document}
\title{\vspace{2cm}\textbf{ On Thermodynamic Stability of Black Holes. Part II: AdS Family of Solutions}\vspace{1cm}}

\author[a]{V. Avramov}
\author[a,b]{H. Dimov}
\author[a]{M. Radomirov}
\author[a,c]{R. C. Rashkov}
\author[a]{T. Vetsov}

\affil[a]{\textit{Department of Physics, Sofia University,}\authorcr\textit{5 J. Bourchier Blvd., 1164 Sofia, Bulgaria}
	
	\vspace{-10pt}\texttt{}\vspace{0.0cm}}

\affil[b]{\textit{The Bogoliubov Laboratory of Theoretical Physics, JINR,}\authorcr\textit{141980 Dubna,
		Moscow region, Russia}
	
	\vspace{-10pt}\texttt{}\vspace{0.0cm}}

\affil[c]{\textit{Institute for Theoretical Physics, Vienna University of Technology,}
  \authorcr\textit{Wiedner Hauptstr. 8–10, 1040 Vienna, Austria}
	\vspace{10pt}\texttt{v.avramov,h\_dimov,radomirov,rash,vetsov@phys.uni-sofia.bg}\vspace{0.1cm}}
\date{}
\maketitle

\begin{abstract}

This study is aimed at providing a thorough analysis of the classical
thermodynamic stability of black holes within the Anti-de Sitter (AdS)
family. We utilize the Nambu bracket formalism to calculate local heat capacities and employ Sylvester's criterion in the mass-energy ensemble to determine both local and global thermodynamic stability regions. Emphasizing the crucial role of the cosmological constant, we establish the conditions necessary for the existence of thermodynamically stable black hole configurations.  Our work highlights the applicability of classical thermodynamics in understanding black hole physics, while acknowledging the potential deviations that may impact the observables of the system beyond the classical level.
\end{abstract}

\thispagestyle{empty}
\tableofcontents

\section{Introduction}


Thermodynamics of black holes \cite{Bekenstein:1972tm, Bekenstein:1973ur, Bekenstein:1974ax, Bekenstein:1975tw, Hawking:1971tu, Hawking:1975vcx, Hawking:1976de, Bardeen:1973gs, Davies:1977bgr, Davies_1978} presents a promising avenue for directly testing our gravitational theories. This stems from the inherent nature of thermal systems to be described by a few observable parameters, a trait which carries over to black holes. These macroscopic parameters include mass, entropy, charge, angular momentum, and additional variables dependent on the underlying gravitational model.

In reality these parameters can be heavily influenced by the surrounding environment. Evidence of such influence arises from the recent discovery of gravitational waves produced by black hole mergers \cite{LIGOScientific:2016aoc} and the apparent images of the supermassive black holes in the center of M87 and our own galaxy \cite{EventHorizonTelescope:2019dse, EventHorizonTelescope:2021dqv, EventHorizonTelescope:2022wkp}. This influence can arise from various factors, such as matter accretion onto the event horizon \cite{Luminet:1979nyg, Page:1974he, Thorne:1974ve, Gyulchev:2019tvk, Gyulchev:2021dvt}, star collisions, or through radiative processes such as Hawking radiation \cite{Hawking:1975vcx}. 

In these scenarios, a natural question arises - under what conditions does the system achieve thermodynamic equilibrium with its surroundings? Despite the well-established criteria for equilibrium in standard thermodynamics \cite{bazarov1964thermodynamics, callen2006thermodynamics, greiner2012thermodynamics, swendsen2020introduction, blundell2010concepts}, this question remains largely unexplored, even for simple black hole configurations. There are only a few instances where generic criteria have been applied to black holes, for instance \cite{Dolan:2014lea, Dolan:2013yca,Sinha21bb,Sinha:2016evq, Sinha:2015zxv, Sinha:2017obh,
Sinha:2020uwf, Sinha19aaa,Avramov:2023eif, Avramov:2023wyi, Peca:1998cs,bardeen:1990a}. 

Our work aims to complement these studies by employing the complete strict  classical criteria for thermodynamic stability of the Anti-de Sitter (AdS) family of black hole solution (see for example \cite{Caldarelli:1999xj} and references therein). The advantage of our approach \cite{Avramov:2023eif, Avramov:2023wyi} is the unified treatment of both local and global thermodynamic stability of such and similar systems \cite{bardeen:1990a, Peca:1998cs, Fernandes:2023byx, Fernandes:2022gjd}. We emphasize the importance of using natural parameters for a given thermodynamic representation and caution against relying solely on non-generic or partial stability criteria.

AdS black holes provide solutions to Einstein's equations in the presence of a negative cosmological constant. They include four primary backgrounds: Schwarzschild-AdS (SAdS), Reissner-Nordstr\"om-AdS (RNAdS), Kerr-AdS, and Kerr-Newman-AdS (KNAdS) black holes. The exploration of these solutions spans a broad field of study, driven by their relevance to diverse disciplines. A prominent example is the well-known AdS/CFT correspondence \cite{Maldacena:1997re}, where AdS black holes are crucial in exploring various aspects of this duality, shedding light on the connection between quantum field theory and gravity at finite temperature. 

Moreover, AdS black holes serve as invaluable tools in exploring phenomena beyond traditional gravitational theory. Their study facilitates investigations into topics like black hole thermodynamics \cite{Hawking:1982dh}, phase transitions in quantum field theory \cite{Witten:1998zw}, and the behavior of strongly correlated systems in condensed matter physics \cite{Hartnoll:2016apf}. The adaptability of AdS black holes broadens their utility to cosmological models \cite{Hebecker:2001nv, Randall:1999ee, Randall:1999vf, Shiromizu:2001jm}, providing frameworks for comprehending the dynamics of the early universe and the implications of negative cosmological constants in gravitational physics. Consequently, a coherent and systematic analysis of their properties, particularly concerning thermodynamic stability, becomes imperative.

This paper constitutes the second part of a series dedicated to investigating the dynamic and thermodynamic characteristics of diverse black holes. Herein, we employ one of the most rigorous criteria for assessing global classical thermodynamic stability, known as the Sylvester criterion, which ensures the positive definiteness of the mass-energy Hessian form. This criterion arises from the strict convexity property of the energy potential in thermodynamics. When extended to black hole systems, classical criteria merely hint at the limitations of classical thermodynamics when considering stability against fluctuations or spontaneous processes. Beyond this realm, quantum laws or deviations from classical setups may also manifest. For instance, certain scenarios suggest potential violations of the third law of thermodynamics by black holes \cite{Davies_1978, Arefeva:2023kpu}, which holds significance in the context of string theory and similar approaches to quantum gravity.

Our analysis of the thermodynamic stability of the AdS family of black hole solutions establishes in details the boundaries and the applicability of classical thermodynamic stability for Anti-de-Sitter black holes, where the cosmological constant is present as a true constant, which is the main goal of this paper. Upon extension of this analysis, a natural question arises: how will classical stability be influenced if the cosmological constant transitions into a thermodynamic parameter \cite{Kastor:2009wy, Dolan:2010ha, Dolan:2011xt, Cvetic:2010jb}? The ramifications of this scenario will be explored in the forthcoming third installment of this  series of works.

Employing a bottom-up approach, we commence with the simplest solutions and proceed towards more intricate configurations. The organization of the text is outlined as follows. In Section \ref{secSAdS}, we explore the thermodynamic stability of the Schwarzschild-AdS black hole, demonstrating how the inclusion of a cosmological constant results in a stable black hole configuration. Moving to Section \ref{secRNAdS}, we analyze the RNAdS scenario, where the presence of an electric charge imposes additional constraints on the regions of thermodynamic stability, albeit maintaining globally stable sectors. Section \ref{secKerrAdS} presents a comprehensive analysis of the thermodynamic stability in the rotating Kerr-AdS black hole, identifying both locally and globally stable regions, along with the identification of possible Davies curves and phase transition points. Finally, in Section \ref{secKNAdS}, we unveil a significantly more intricate structure in the thermodynamic phase space of the KNAdS black hole. Here, distinct Davies surfaces correspond to the divergences of heat capacities, alongside higher-dimensional regions of local and global thermodynamic stability. Additionally, in Appendix \ref{secB}, we illustrate how the SAdS, RNAdS and Kerr-AdS thermodynamic stability is embedded within the KNAdS state space.

\section{Thermodynamic stability of Schwarzschild-AdS black hole}\label{secSAdS}

The metric of the Schwarzschild-anti-de-Sitter black hole solution (SAdS), with negative cosmological constant\footnote{We will adopt units for which $\kappa = 8\pi$, thus $G = 1$.} $\Lambda=-3/l^2$, is written by\footnote{In static spherical coordinates. In the limit $l\to \infty$ one recovers the Schwarzschild case.} \cite{Hawking:1982dh, Witten:1998zw}:
\begin{equation}
   ds^2= - \left(1-\frac{2M}{r}+\frac{r^2}{l^2}\right)dt^2 +\bigg( 1-\frac{2M}{r}+\frac{r^2}{l^2} \bigg)^{\!-1}  dr^2 +r^2 \big(d\theta ^2+\sin^2\!\theta\, d\phi ^2 \big),
\end{equation}
where $l>0$ is the radius of the AdS space and $M$ is the mass of the black hole.
The event horizon is the real solution to the cubic equation $r^3 +l^2r -2l^2 M=0$ with $M>0$ and $l>0$.  The thermodynamics of the SAdS black hole is defined by its mass $M$, entropy $S$ and temperature $T$, given by
\begin{equation}
    M=\frac{1}{2}\sqrt{\frac{S}{\pi}} \bigg(1+\frac{S}{\pi l^2}\bigg), \quad T=\frac{\partial M}{\partial S}=\frac{1}{4  \sqrt{\pi S}} \bigg(1+\frac{3 S}{\pi  l^2}\bigg),\quad dM=T dS.
\end{equation}
It will be useful to introduce the following rescaled parameters:
\begin{equation}\label{RescaledParSAdS}
m=\frac{2M}{l}, \quad \tau=2\pi l T, \quad  s=\frac{S}{\pi l^2},
\end{equation}
hence one can write:
\begin{equation}
m=\sqrt{s\,} (1 +s),\quad \tau= \frac{\partial m}{\partial s}=\frac{1 +3s}{2\sqrt{s\,}}, \quad dm=\tau ds.
\end{equation}

In the mass-energy representation $m=m(s)$ there is only one condition for classical global thermodynamic stability, which  naturaly follows from the convexity of the mass potential:
\begin{equation}\label{eqGTDSSchAds}
  \mathcal{H}_{ss}\equiv  \frac{\partial^2 m}{\partial s^2} =\frac{3s -1}{4s^{3/2}} > 0\quad\Rightarrow \quad s>\frac{1}{3}.
\end{equation}
This suggests that the SAdS solution can be thermodynamically stable only if the black hole has entropy  $S>\pi l^2/3$. On the other hand, the heat capacity of the Schwarzschild-AdS black hole,
\begin{equation}\label{SAdSHeatC}
   C=\frac{\partial m}{\partial \tau}= \frac{\partial m}{\partial s} \frac{1}{\frac{\partial \tau}{\partial s}} =\frac{2 s (1 +3s)}{3s -1}>0\quad \Rightarrow\quad s>\frac{1}{3},
\end{equation}
determines its local thermodynamic stability. 
It is evident that both conditions for global and local thermodynamic equilibrium coincide for  $S>\pi l^2/3$.  Therefore, one can consider the Schwarzschild-AdS solution to be locally and globally stable from classical thermodynamic standpoint, which  is in contrast to the Schwarzschild case ($l\to \infty$) \cite{Avramov:2023eif}. Hence, the introduction of a finite cosmological parameter $l$ leads to thermodynamically stable black hole configuration.  This, however, is not a trivial result. For $S<\pi  l^2/3$, the Schwarzschild-AdS black hole is classically unstable from thermodynamic standpoint, thus can radiate. On the special point $S=\pi  l^2/3$ the heat capacity diverges, thus defining a Davies phase transition point.

A note of caution should be marked here. If one considers positive semi-definiteness of the Hessian $\mathcal{H}_{ss}\geq 0$ there will be a contradiction between the local and the global thermodynamic stability, since the point $S=\pi l^2/3$ will be a part of the global region of stability. However, it is shown that it corresponds to a divergence in the heat capacity. This is an explicit example of why one should use the strong positive definiteness of the Hessian $\mathcal{H}_{ss}>0$ instead of the weaker semi-definite one.

The explanation of the stability regions in the context of Schwarzschild Anti-de Sitter (SAdS) is as follows. In the case of stable black hole configurations, the criterion $S =\pi r_+^2> \pi l^2/3$ implies $r_+^2 > l^2/3$, indicating that the black hole event horizon is comparable to the size of the AdS space. On the other hand, for unstable configurations, $r_+^2 < l^2/3$, suggesting that observable astrophysical black holes with significantly smaller radii will emit radiation. The critical radius $r_+^2=l^2/3$ is the Davies point separating thermodynamically stable and unstable regions, i.e it represents the local minimum of the Hawking temperature,
\begin{equation}
    T_{min}=\frac{\sqrt{3}}{2\pi l}.
\end{equation}

\section{Thermodynamic stability of Reissner-Nordstr\"om-AdS black hole}\label{secRNAdS}

Next we investigate the Reissner-Nordström-AdS black hole solution, where the inclusion of an electric charge imposes supplementary constraints on the regions of thermodynamic stability. Our analysis discerns both local and global stable regions, and also includes potential Davies curves and phase transition points. 

\subsection{The Reissner-Nordstr\"om-AdS black hole thermodynamics}

The metric of the Reissner-Nordstr\"om-AdS (RNAdS) black hole is written by \cite{Chamblin:1999tk, Chamblin:1999hg,Louko:1996dw}:
\begin{equation}
   ds^2= - \left( 1-\frac{2M}{r} +\frac{Q^2}{r^2} +\frac{r^2}{l^2}\right)dt^2 +\left(1-\frac{2M}{r} +\frac{Q^2}{r^2} +\frac{r^2}{l^2}\right)^{\!-1} dr^2 +r^2 \big(d\theta ^2+\sin ^2\theta d\phi ^2 \big),
\end{equation}
where $M$ is the mass and $Q\in \mathbb{R}$ is the electric charge of the RNAdS. 
The event horizon $r_+$ is a solution to the quartic equation $r^4 + l^2r^2 - 2l^2Mr + l^2Q^2 = 0$. This equation also provides the inner Cauchy horizon $r_-$. The existence of a black hole is established by the condition $r_+ > r_-$, which is related to the critical mass parameter $M_*$ of the extremal RNAdS ($r_+ = r_{-}$) \cite{Louko:1996dw}:
\begin{equation}\label{eqStrongerECRNAdS}
   M>M_*= \frac{l}{3 \sqrt{6}}\left(\sqrt{\frac{12 Q^2}{l^2}+1}+2\right) \bigg(\sqrt{\frac{12 Q^2}{l^2}+1}-1\bigg)^{1/2}.
\end{equation}

\begin{figure}[H]
\centering
\includegraphics[scale=0.5]{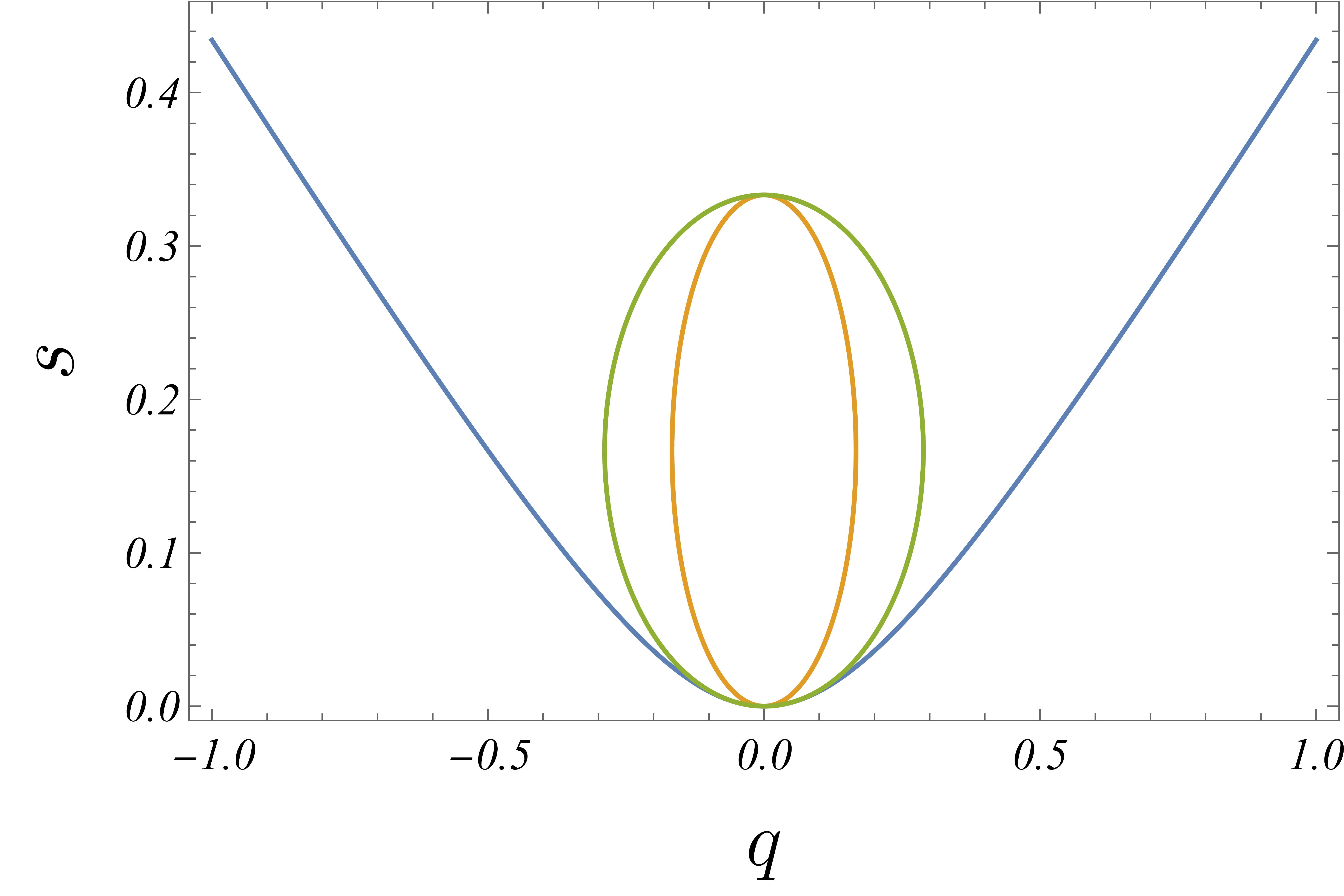}
\caption{ The existence of the black hole is defined above the blue hyperbola $(b>0)$. Above the blue hyperbola and outside the green ellipse $(b>0,\,g>0)$ the RNAdS black hole can be globally stable. Inside the green ellipse $(g<0)$  the black hole is unstable for processes with constant $\phi$.  Inside the orange ellipse $(o<0)$ the black hole is unstable for processes with constant $q$.}
\label{RNAdS}
\end{figure}

Similarly to the SAdS case we can introduce the following rescaled  parameters\footnote{Our scales differ from those in \cite{Banerjee:2010da} by some factors of $\pi$ or 2.}:
\begin{equation}\label{RescaledParRNAdS}
m=\frac{2M}{l}, \quad \tau=2\pi l T, \quad  s=\frac{S}{\pi l^2}, \quad \phi=2\Phi, \quad q=\frac{Q}{l},
\end{equation}
where $T$ is the Hawking temperature, $S$ is the entropy and $\Phi$ is the electric potential of the solution.
In these notations the thermodynamics of the RNAdS black hole takes the form:
\begin{align}\label{eqRNAdSParams}
   m= \frac{s^2 + s+ q^2}{ \sqrt{s}},\quad \tau=\frac{\partial m}{\partial s}\bigg|_q=\frac{3 s^2+ s- q^2}{2s^{3/2}}, \quad \phi=\frac{\partial m}{\partial q}\bigg|_{s}=\frac{2q}{\sqrt{s}}, \quad dm=\tau ds +\phi dq.
\end{align}
Here we assume the energy representation, hence  $m$, $\tau$ and $\phi$ become functions of $(s,q)$. Imposing positive entropy $s>0$ and positive temperature $\tau>0$, we find 
\begin{equation}\label{eqRNAdSExistQ}
q^2<s(1+3s).
\end{equation}
This condition coincides with (\ref{eqStrongerECRNAdS}), which makes (\ref{eqRNAdSExistQ})  sufficient for the existence of the RNAdS solution. Therefore the physically relevant region is above the blue hyperbola in ($s,q$) space (Fig.\,\ref{RNAdS}) given by the following equation
\begin{equation}\label{BlueHyperbola}
3s^2+s-q^2=0.
\end{equation}
We can write the existence condition \eqref{eqRNAdSExistQ} together with \eqref{eqRNAdSParams} also as:
\begin{equation}\label{existrelations}
(1+s)\sqrt{s}<m<2(1+2s)\sqrt{s}, \quad 0<\tau<\frac{1+3s}{2\sqrt{s}}, \quad |\phi|<2\sqrt{1+3s}.
\end{equation}

\subsection{Global thermodynamic stability of RNAdS black hole}

We can now study the conditions for global thermodynamic stability of the RNAdS black hole. 
For this purpose it would be useful to define the following functions:
\begin{align}
&b(s,q)=3s^2+s-q^2, \\
\label{GreenEllipse}
&g(s,q)=3s^2-s+q^2, \\
\label{OrangeEllipcse}
&o(s,q)=3s^2-s+3q^2,
\end{align}
where the chosen letters ($b$-blue, $g$-green, $o$-orange) correspond to the colors of the curves depicted on Fig \ref{RNAdS}. The Hessian of the mass in $(s,q)$ space is given by
\begin{equation}
\mathcal{H}=\left(\!\begin{array}{cc}
\mathcal{H}_{ss} & \mathcal{H}_{sq}\\[5pt]
\mathcal{H}_{qs} & \mathcal{H}_{qq}\\
\end{array}\!\right)=
\left(\!
\begin{array}{ccc}
\frac{\partial^2 m}{\partial s^2} & \frac{\partial^2 m}{\partial s \partial q}\\[5pt]
\frac{\partial^2 m}{\partial q\partial s} & \frac{\partial^2 m}{\partial q^2}
\end{array}
\!\right)
=\left(\!
\begin{array}{cc}
 \frac{o(s,q)}{4 s^{5/2}} & -\,\frac{q}{ s^{3/2}} \\[5pt]
 -\,\frac{q}{ s^{3/2}} & \frac{2}{\sqrt{s}} \\
\end{array}
\!\right).
\end{equation}
The conditions for global thermodynamic stability require positive definite Hessian, namely
\begin{align}
&\mathcal{H}_{ss}=\frac{o(s,q)}{4 s^{5/2}}> 0, \quad \mathcal{H}_{qq}=\frac{2}{\sqrt{s\,}}> 0,\quad \det\mathcal{H}=\frac{g(s,q)}{2 s^3}> 0.
\end{align}
One notes that $\mathcal{H}_{qq}$ is always positive, while $\mathcal{H}_{ss}$ is positive only if 
\begin{equation}\label{orangeEllipse}
o(s,q)>0.
\end{equation}
This condition is satisfied outside the orange ellipse, given by $o(s,q)=0$ (Fig, \ref{RNAdS}), which  can be identified as the Davies critical curve for $C_q$ (see Eq.\eqref{Cq}).
The determinant of the Hessian is positive if
\begin{equation}\label{greenEllipse}
g(s,q)>0,
\end{equation}
which is fulfilled outside the green ellipse, defined by $g(s,q)=0$. The latter corresponds to the Davies curve for $C_\phi$ \eqref{Cfi}.
The conditions for global thermodynamic stability should agree with the condition for the existence (\ref{eqRNAdSExistQ}) of the black hole. This leads to
\begin{equation}\label{GTS_RN}
s(1-3s)<q^2<s(1+3s),
\end{equation}
which resides in the region above the blue hyperbola $(b>0)$ and outside the green ellipse $(g>0)$. The equation \eqref{GTS_RN} defines the parametric region that characterizes the global thermodynamic stability of the RNAdS solution. Understanding the significance of this region, we can visualize the macrostate of the black hole as a point within the $(s, q)$ space. This point can either move within the $(s, q)$ space through certain processes or remain fixed, maintaining equilibrium with its surroundings. However, outside the region of global stability, the point cannot remain fixed, and the black hole will undergo spontaneous changes in its macrostate, preventing it from reaching equilibrium.

\subsection{Local thermodynamic stability of RNAdS black hole}

The relevant heat capacities of the RNAdS black hole in $(s,q)$ space are given by\footnote{Local thermodynamic stability of Reissner-Nordstr\"om-anti-de Sitter black hole has been investigated in a similar approach by \cite{Peca:1998cs}.}
\begin{align}
&C_m=\tau \frac{\partial s}{\partial \tau}\bigg|_{m}=\tau\frac{ \{s,m\}_{s,q}}{\{\tau,m\}_{s,q}}= \frac{s \left(3 s^2 +s -q^2\right)}{3s^2 +q^2},
    \\ \label{Cfi}
&C_{\phi }=\tau \frac{\partial s}{\partial \tau} \bigg|_{\phi} =\tau\frac{ \{s,\phi \}_{s,q}}{\{\tau,\phi \}_{s,q}}= \frac{2s \left(3s^2 +s -q^2\right)}{3s^2 -s +q^2},
    \\ \label{Cq}
&C_q=\tau \frac{\partial s}{\partial \tau} \bigg|_{q}=\tau\frac{ \{s,q\}_{s,q}}{\{\tau,q\}_{s,q}}
    =\frac{2 s \left(3 s^2 +s -q^2\right)}{3s^2-s+3q^2}.
\end{align}

Let us consider a processes with constant mass. One can see that $C_m$ is always positive in the region of existence (\ref{eqRNAdSExistQ}), thus the black hole is locally stable there with respect to $C_m$. Furthermore, outside the green ellipse $(g>0)$ (Fig. \!\ref{RNAdS}) the black hole can be globally stable. However, inside the green ellipse $(g<0)$ the equilibrium is only local with respect to processes with constant mass.

Next, we consider processes with constant electric potential. The condition $C_\phi>0$ matches the region of global stability (\ref{GTS_RN}), hence above the blue hyperbola $(b>0)$ and outside the green ellipse $(g>0)$ the RNAdS black hole can be locally and globally stable. Inside the green ellipse $C_\phi<0$ and the black hole is  thermodynamically unstable. Note that the green ellipse $(g=0)$ is a Davies curve for $C_\phi$, thus defining the locus of possible phase transitions.  

Finally, we consider processes with constant electric charge. In this case,  the
region of local stability $C_q>0$ is located outside the orange ellipse $(o>0)$. In the sector outside the green ellipse $(g>0)$ the black hole can be locally and globally stable. Between the two ellipses $(o>0,\,g<0)$ the black hole could maintain only a local equilibrium. Inside the orange ellipse the black hole is unstable $C_q<0$. Note that the orange ellipse is Davies curve for $C_q$, representing possible phase transitions.

All of the heat capacities are positive in the region of global stability \eqref{GTS_RN}, which is defined above the blue hyperbola $(b>0)$ and outside the green ellipse $(g>0)$ (Fig. \ref{RNAdS}). However, we showed that the RNAdS solution also admits some partial regions of local thermodynamic stability, located inside the green ellipse. Once again global stability is assured due to the presence of a cosmological constant.

\subsection{Davies curves and phase transitions of RNAdS}

The critical Davies curves mark the points where heat capacities exhibit divergences or change signs ($C\to 0, \pm \infty$). Specifically, for $C_m$, one has:
\begin{align}
    &C_m\to
     +0 \quad \text{for} \quad b(s,q)\to +0, \quad  (\tau\to +0),
\end{align}
It is important to note that the limit above is not direct. This is due to the fact that at fixed mass the parameters $s$ and $q$ are not independent of each other, i.e. they satisfy \eqref{eqRNAdSParams} 
\begin{equation}
m=const= \frac{s^2 + s+ q^2}{ \sqrt{s}} \to 2(1+2s)\sqrt{s}, \quad {\text{when}} \quad b(s,q)\to +0.
\end{equation}
However, one notes that this particular value of the mass corresponds to the upper limit from  \eqref{existrelations}. Hence the limit $b(s,q)\to +0$ indicates a true phase transition point. The curve $b(s,q)=0$ corresponds to the blue hyperbola as shown on Figure \ref{RNAdS}. Moreover, examining the limit $q = 0$ presents no issue, as both mass and heat capacity remain finite and regular:
\begin{equation}
m=(1+s)\sqrt{s}, \quad C_m= \frac{1+3s}{3},
\end{equation}
which corresponds to the SAdS scenario, albeit with a different heat capacity $C_m$ compared to \eqref{SAdSHeatC}. This discrepancy arises from our approach to this point via a specific process with an associated heat capacity $C_m$, influenced by the presence of a charge. In the absence of any charge ($q = 0$), the system can indefinitely persist in the SAdS state, characterized by its distinct heat capacity as defined in \eqref{SAdSHeatC}. Consequently, we can relax the existence conditions \eqref{existrelations} to encompass the lower limit of mass and the upper limit of temperature:
\begin{equation}
(1+s)\sqrt{s}\leq m<2(1+2s)\sqrt{s}, \qquad 0<\tau \leq \frac{1+3s}{2\sqrt{s}}.
\end{equation}

Similar analysis can be done on the remaining heat capacities. For example, considering a process with constant electric potential, one has   
\begin{align}
&C_\phi\to\left\{
\begin{array}{l}
+0,\,\,\, b(s,q)\to +0, \quad  (\tau\to +0),\\
+\infty,\,\,\, g(s,q)\to +0,\\
 -\infty,\,\,\, g(s,q)\to -0.
\end{array} 
\right. 
\end{align}
Consequently, considering processes with constant charge, one finds
\begin{align}
&C_q\to\left\{
\begin{array}{l}
+0,\,\,\,b(s,q)\to +0, \quad  (\tau\to +0),\\
+\infty,\,\,\, o(s,q)\to +0,\\
 -\infty,\,\,\, o(s,q)\to -0.
\end{array} 
\right.
\end{align}
All heat capacities converge to zero along the blue hyperbola (\ref{BlueHyperbola}), where the black hole temperature approaches absolute zero, $\tau\to0$. This signifies the extremal scenario\footnote{It's worth noting that extremal cases may hold significance in other theories such as modified gravity or string theory, where violations of the third law of thermodynamics can occur. See for instance \cite{Davies:1977bgr,Arefeva:2023kpu}.}, which remains unattainable through a finite number of processes due to the third law of thermodynamics. Moreover, $C_\phi\to +\infty$ as we approach the outer boundary of the green ellipse $(g\to +0)$, and to negative infinity when approaching from the inside $(g\to -0)$. Additionally, $C_q$ diverges towards positive (negative) infinity when approaching the outer (inner) edge of the orange ellipse.

\section{Thermodynamic stability of Kerr-AdS black hole}\label{secKerrAdS}

In this section, we provide an in-depth analysis of the thermodynamic stability of Kerr-AdS black hole equipped with an angular momentum. We identify both locally and globally stable regions, as well as highlight possible Davies curves and phase transition points.

\subsection{Kerr-AdS thermodynamics}
The Kerr-AdS metric is given by (\ref{eqKNAdSMetric}) with the charge parameter $\mathcal{Q}=0$, \cite{Gibbons:2004ai, Carter:1968ks}. Its thermodynamics can be written in the form:
\begin{align}\label{eqKAdSParams}
m=\frac{\sqrt{(1+s)(j^2+s^2+s^3)}}{\sqrt{s\,}},\,\,\, \tau=\frac{s^2(1+s)(1+3s) -j^2}{2\sqrt{s^3(1+s) (j^2+s^2+s^3)}},   \,\,\, \omega=\frac{j\sqrt{1+s}}{\sqrt{s(j^2+s^2+s^3)}},
\end{align}
where once again we have the following rescaled parameters:
\begin{equation}\label{RescaledPar1}
m=\frac{2M}{l}, \quad \tau=2\pi l T, \quad  s=\frac{S}{\pi l^2}, \quad \omega= l \Omega, \quad j=\frac{2J}{l^2}, \quad dm=\tau ds +\omega dj.
\end{equation}
In this case, the mass $m$, the temperature $\tau$ and the angular velocity $\omega$ are functions of the entropy $s$ and the angular momentum $j$. It would be useful to define the following functions:
\begin{align}\label{BlueCurve}
&\mathfrak{b}(s,j)= s^2(1+s)(1+3s) -j^2,\\
\label{KerrGreen}
&\mathfrak{g}(s,j)= s^2 (1+s)\big(2 \sqrt{s (1+s)}-(1+s)\big) -j^2,\\
\label{KerrHsseq0}
&\mathfrak{o}(s,j)=s^4 (1+s)^3 (3s-1) +6j^2s^2 (1+s)^2 (1+2s) +j^4 (3+4s).
\end{align}

\begin{figure}[H]
\centering
\includegraphics[scale=0.5]{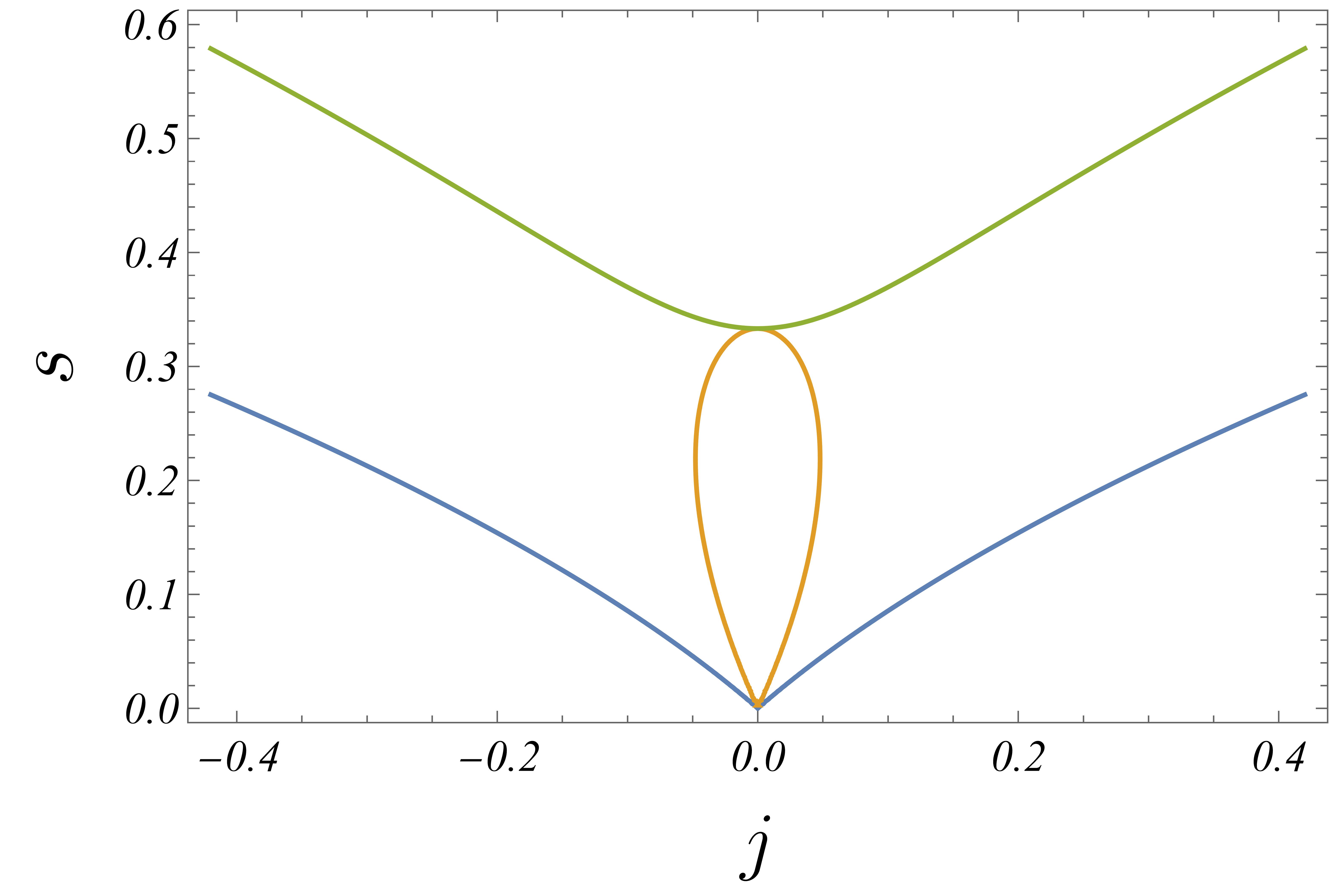}
\caption{ The existence of the black hole is defined above the blue curve $(\mathfrak b>0)$. Above the green curve $(\mathfrak g>0)$ the system can be in global equilibrium.  Below the green curve $(\mathfrak g<0)$ the black hole is unstable for processes with constant $\omega$.  Inside the orange curve ($\mathfrak 0<0$) the black hole is unstable for processes with constant $j$. }\label{KerrAdS}
\end{figure}

The region of existence of Kerr-AdS black hole in ($s,j$) space is defined by $s>0$ and $\tau>0$, which lead to the relation
\begin{equation}\label{eqKerrAdSExist}
j^2<s^2(1+s)(1+3s).
\end{equation}
It is satisfied above the blue curve $\mathfrak{b}(s,j)=0$ (Fig.\,\ref{KerrAdS}). The condition for existence \eqref{eqKerrAdSExist} together with \eqref{eqKAdSParams} lead to the following restrictions on the parameters of the black hole:
\begin{equation}
(1+s)\sqrt{s}\leq m<(1+s)\sqrt{s(2+3s)}, \quad 0<\tau\leq \frac{1+3s}{2\sqrt{s}}, \quad |\omega|<\sqrt{\frac{(1+s)(1+3s)}{s(2+3s)}}.
\end{equation}

\subsection{Global thermodynamic stability of Kerr-AdS black hole}

The Hessian of the mass in $(s,j)$ space yields\footnote{See \cite{Monteiro:2009ke} for grand canonical treatment of stability.}
\begin{equation}
\mathcal{H}=\left(\!\begin{array}{cc}
\mathcal{H}_{ss} & \mathcal{H}_{sj} \\[5pt]
\mathcal{H}_{js} & \mathcal{H}_{jj} \\
\end{array}\!\right)=
\left(\!
\begin{array}{ccc}
\frac{\partial^2 m}{\partial s^2} & \frac{\partial^2 m}{\partial s \partial j}\\[5pt]
\frac{\partial^2 m}{\partial j\partial s} & \frac{\partial^2 m}{\partial j^2} \\
\end{array}
\!\right),
\end{equation}
with components given by:
\begin{align}\label{KerrH_ss}
&\mathcal{H}_{ss}=\frac{\mathfrak{o}(s,j)}{4 \sqrt{s^5 (1+s)^3 \left(j^2+s^2+s^3\right)^3 }},\quad \mathcal{H}_{jj}= \left(\frac{s (1 +s)}{ j^2+s^2+s^3}\right)^{3/2}, \\
&\mathcal{H}_{sj}= \mathcal{H}_{js}= -\,\frac{j \big( j^2+3 s^2 (1+s)^2 \big) }{2\sqrt{s^3 (1+s) \left(j^2+s^2+s^3\right)^3\,} }, \quad 
\det \mathcal{H}=\frac{\mathfrak{g}(s,j)} {4s^3 (1+s) \left( j^2+s^2+s^3\right)^2}.
\end{align}
The conditions for global thermodynamic stability require
\begin{equation}
\mathcal{H}_{ss}>0, \quad \mathcal{H}_{jj}>0, \quad \det\mathcal{H}>0.
\end{equation}
One notes that $\mathcal{H}_{jj}$ is always positive, while $\mathcal{H}_{ss}$ is positive when the numerator is positive, i.e.
\begin{equation}\label{KerrHss>0}
\mathfrak{o}(s,j)>0.
\end{equation}
This relation is fulfilled outside the closed orange curve, defined by $\mathfrak{o}(s,j)=0$ (Fig.\,\ref{KerrAdS}).
Furthermore, the determinant of the Hessian $\det\mathcal{H}$ is positive when its numerator is positive,
\begin{equation}\label{KerrGlStab}
\mathfrak{g}(s,j)>0. 
\end{equation}
This condition is satisfied above the green curve, given by $\mathfrak{g}(s,j)=0$.
Relations (\ref{KerrHss>0}) and (\ref{KerrGlStab}) indicate that the Kerr-AdS black hole can reside in global equilibrium above the green curve $(\mathfrak{g}>0)$, i.e.
\begin{equation}\label{KerrGlStab1}
j^2< s^2 (1+s)\big(2 \sqrt{s (1+s)}-(1+s)\big).
\end{equation}
Evidently, when $j=0$ one recovers $s>1/3$ from Eq. (\ref{eqGTDSSchAds}) for the SAdS case. This is a regular limit and not a phase transition point.

\subsection{Local thermodynamic stability of  Kerr-AdS black hole}

The local heat capacities of the Kerr-AdS black hole in $(s,j)$ space are given by
\begin{align}
&C_m=\tau \frac{\partial s}{\partial \tau}\bigg|_{m}=\tau\frac{ \{s,m\}_{s,j}}{\{\tau,m\}_{s,j}}= \frac{s (1+s)\, \mathfrak{b}(s,j)}{s^2 (1+s)^2 (1+6s) +j^2 (1+2s)},
    \\
&C_{\omega }=\tau \frac{\partial s}{\partial \tau} \bigg|_{\omega} =\tau\frac{ \{s,\omega \}_{s,j}}{\{\tau,\omega \}_{s,j}}= \frac{2 s^3 (1+s)^2\, \mathfrak{b}(s,j)}{s^4 (1+s)^3 (3s-1) -2j^2 s^2 (1+s)^2 -j^4},
    \\
&C_j=\tau \frac{\partial s}{\partial \tau} \bigg|_{j}=\tau\frac{ \{s,j\}_{s,j}}{\{\tau,j\}_{s,j}}
    =\frac{2 s (1+s) (j^2 +s^2 +s^3)\, \mathfrak{b}(s,j)}{\mathfrak{o}(s,j)}.
\end{align}

In the region of existence (\ref{eqKerrAdSExist})  of the black hole $C_m>0$, hence we have local stability with respect to processes with constant mass. Above the green curve one may have local and global equilibrium altogether (Fig. \ref{KerrAdS}). Between the blue and the green curves the black hole is only in local equilibrium. 

Under a process with constant angular velocity one has $C_{\omega}>0$ within the region of global stability (\ref{KerrGlStab1}), given above the green curve. Between the blue and the green curves the black hole is unstable due to $C_\omega<0$.

Finally, we consider processes with constant angular momentum. In this case,  the
region of local stability $C_j>0$ is located outside the closed orange curve $(\mathfrak o>0)$. In the sector above the green curve $(\mathfrak g>0)$ the black hole can be locally and globally stable. Between the blue and the green curve and outside the closed orange curve ($\mathfrak b>0,\,\mathfrak g<0$ and $\mathfrak o>0)$ the black hole could maintain only a local equilibrium. Inside the closed orange curve $(\mathfrak o<0)$ the black hole is unstable due to $C_j<0$. 
 
Finally, one notes that all of the heat capacities are positive in the region of global stability (above the green curve). 

\subsection{Critical points and phase transitions of   Kerr-AdS black hole}

The critical curves of the Kerr-AdS black hole are defined by:
\begin{align}
C_m\to
    0, \,\, \mathfrak{b}(s,j)\to +0,
\quad   C_j\to\left\{
\begin{array}{l}\!\!
    0, \,\, \mathfrak{b}(s,j)\to +0,\\
    \!\!\infty, \,\, \mathfrak{g}(s,j)\to +0, \\
   \!\! -\infty, \,\, \mathfrak{g}(s,j)\to -0,
\end{array} 
\right.\quad     C_\omega \to\left\{
\begin{array}{l}
\!\!0, \,\, \mathfrak{b}(s,j)\to +0,\\
\!\!\infty, \,\, \mathfrak{o}(s,j)\to +0, \\
\!\!-\infty, \,\, \mathfrak{o}(s,j)\to -0 .
\end{array} 
\right.
\end{align}
All of the heat capacities go to zero on the blue curve (\ref{BlueCurve}), corresponding to the extremal case. Furthermore, $C_\omega\to \infty (-\infty)$ when we approach the green curve from above (below). Additionally, $C_j\to \infty (-\infty)$ when we approach the closed orange curve (\ref{KerrHsseq0}) from the outside (inside). Finally the limit $j\to 0$ is a regular limit towards the SAdS black hole. 

\section{Thermodynamic stability of Kerr-Newman-AdS black hole}\label{secKNAdS}

Finally, we can study the classical thermodynamic stability of the  Kerr-Newman-AdS black hole (KNAdS).

\subsection{Thermodynamics in Kerr-Newman-AdS spacetime}

The metric of the Kerr-Newman-AdS black hole\footnote{It reduces to the standard KN solution for $l\to \infty$.} is given by \cite{Carter:1968ks, Plebanski:1976gy, Caldarelli:1999xj}:
\begin{equation}\label{eqKNAdSMetric}
    ds^2=-\frac{\Delta_r}{\rho^2}\bigg(dt-\frac{a \sin^2\theta}{\Xi_-}d\phi\bigg)^2+\frac{\rho^2}{\Delta_r} dr^2+\frac{\rho^2}{\Delta_\theta} d\theta^2+\frac{\Delta_\theta \sin^2 \theta}{\rho^2}\bigg(a dt-\frac{r^2+a^2}{\Xi_-} d\phi\bigg)^2,
\end{equation}
where the following notations have been adopted:
\begin{align}\label{eqKNAdSmetric}
    & \rho^2=r^2+a^2 \cos^2\theta,\quad \Xi_{-}=1-\frac{a^2}{l^2},\quad \Xi_{+}=1+\frac{a^2}{l^2},
    \\
    &\Delta_r=(r^2+a^2) \bigg(1+\frac{r^2}{l^2}\bigg)-2 \mu r+\mathcal{Q}^2,\quad \Delta_\theta=1-\frac{a^2}{l^2} \cos^2 \theta.
\end{align}
Here $a$ is the specific angular momentum, $\mathcal{Q}$ is the specific charge parameter, $\mu$ is the specific mass parameter as defined in Eq. \eqref{eqMSJQ}. Additionally, $l$ is the AdS radius  related to the cosmological constant by $\Lambda=-3/l^2$.
For $a^2<l^2$ and $\mu>\mu_*$ the metric (\ref{eqKNAdSmetric})
represents AdS black hole with an event horizon located at $r = r_+$, where $r_+$ is the largest solution of $\Delta_r=0$.  The parameter $\mu_*$ is the critical mass parameter defined by the extremal case $r_+=r_-$ \cite{Caldarelli:1999xj}:
\begin{equation}\label{eqMcrit}
  \mu_*=  \frac{l}{3 \sqrt 6}\bigg(\sqrt{\Xi_+^2+\frac{12}{l^2}(a^2+\mathcal{Q}^2)}+2\Xi_+\bigg)
    \bigg(\sqrt{\Xi_+^2+\frac{12}{l^2}(a^2+\mathcal{Q}^2)}-\Xi_+\bigg)^{1/2}.
\end{equation}

The thermodynamics of the KNAdS black hole in mass-energy representation has been presented in \cite{Caldarelli:1999xj}. In this case, the mass $M$, the entropy $S$, the angular momentum $J$, and the charge $Q$ of the solution can be written by
\begin{equation}\label{eqMSJQ}
    M=\frac{\mu}{\Xi_{-}^2},\quad S=\frac{\pi (r^2_{+}+a^2)}{\Xi_-}, \quad    J=\frac{a \mu}{\Xi_{-}^2},\quad Q=\frac{\mathcal{Q}}{\Xi_{-}}.
\end{equation}
The Hawking temperature $T$, the angular velocity $\Omega$ and the charge potential $\Phi$ yield:
\begin{equation}\label{eqKNAdSTOmP}
    T=\frac{r_{+} }{{4 \pi (r^2_{+}+a^2)}}\bigg(\Xi_{+}+\frac{3 r_{+}^2}{l^2}-\frac{a^2+\mathcal{Q}^2}{r_{+}^2}\bigg),\quad \Omega=\frac{a}{{r^2_{+}+a^2}} \bigg(1+\frac{r^2_{+}}{l^2}\bigg), \quad \Phi=\frac{\mathcal{Q}\, r_+}{r_{+}^2+a^2}.
\end{equation}
The fundamental relation $M = M(S, J, Q)$ can be obtained by solving Eqs. (\ref{eqMSJQ}) with respect to $r_+, a, \mu$ and $\mathcal{Q}$, i.e.
\begin{equation}\label{eqEVHKNADS}
   r_+= \frac{\sqrt{l^2 M^2 S-J^2 \left(\pi  l^2+S\right)}}{\sqrt{\pi } l M},\quad a= \frac{J}{M},\quad \mu= \frac{\left(J^2-l^2 M^2\right)^2}{l^4 M^3},\quad \mathcal{Q}= Q\bigg(1-\frac{J^2}{l^2 M^2}\bigg).
\end{equation}
Inserting these expressions back into Eqs. (\ref{eqMSJQ}) and (\ref{eqKNAdSTOmP}) one obtains\footnote{For the extended phase space thermodynamics see \cite{Caldarelli:1999xj}, where $\ell^2=-3/\Lambda$, $P=-\Lambda/(8 \pi)$, $\Theta=-V/(8\pi)$ and the mass of the black hole $M=E+P V$ is now identified as the enthalpy of spacetime. See also \cite{Imseis:2020vsw}.} the mass-energy thermodynamic representation of KNAdS system:
\begin{align}
    &M^2=\frac{S}{4 \pi}+\frac{\pi}{4 S} (4 J^2+Q^4)+\frac{Q^2}{2}+\frac{J^2}{l^2}+\frac{S}{2 \pi l^2}\bigg(Q^2+\frac{S}{\pi}+\frac{S^2}{2 \pi^2 l^2}\bigg),
\\
    &T=\frac{\partial M}{\partial S}\bigg|_{J,Q}=\frac{1}{8 \pi M} \bigg[1-\frac{\pi^2}{S^2} (4 J^2+Q^4)+\frac{2}{l^2}\bigg(Q^2+\frac{2 S}{\pi}\bigg)+\frac{3 S^2}{\pi^2 l^4}\bigg],
    \\
    &\Omega=\frac{\partial M}{\partial J}\bigg|_{S,Q}=\frac{\pi J}{M S}\bigg(1+\frac{S}{\pi l^2}\bigg),
    \\
    &\Phi=\frac{\partial M}{\partial Q}\bigg|_{S,J}=\frac{\pi Q}{2 M S}\bigg(Q^2+\frac{S}{\pi}+\frac{S^2}{\pi^2 l^2}\bigg).
\end{align}
Consequently, the first law for KNAdS takes the form
\begin{equation}\label{eqFLTKNADS}
    dM=T dS+\Omega dJ+\Phi dQ.
\end{equation}
Assuming $l>0$ and $l\neq \infty$ we can rescale the thermodynamic parameters:
\begin{equation}\label{RescaledPar}
m=\frac{2M}{l}, \quad \tau=2\pi l T, \quad  s=\frac{S}{\pi l^2}, \quad \omega= l \Omega, \quad j=\frac{2J}{l^2}, \quad \phi=2\Phi, \quad q=\frac{Q}{l} ,
\end{equation}
hence the following form of the KNAdS thermodynamics:
\begin{align}\label{eqm}
dm&=\tau ds +\omega dj +\phi dq, \nonumber \\
m&= \frac{\sqrt{j^2(1+s) +(s^2+s+q^2)^2}}{\sqrt{s\,}}, \\
\tau&=\frac{\partial m}{\partial s}\bigg|_{j,q}=\frac{(s^2+s+q^2)(3s^2+s-q^2)-j^2}{2s^{3/2}\sqrt{j^2(1+s)+(s^2+s+q^2)^2}}, \\
\omega&=\frac{\partial m}{\partial j}\bigg|_{s,q}= \frac{j(1+s)}{\sqrt{s\,}\sqrt{j^2(1+s)+(s^2+s+q^2)^2}}, \\\label{eqPhi}
\phi&=\frac{\partial m}{\partial q}\bigg|_{s,j}= \frac{2q(s^2+s+q^2)}{\sqrt{s\,}\sqrt{j^2(1+s)+(s^2+s+q^2)^2}}.
\end{align}

Let us look at the possible restrictions on the parameters. For $\tau>0$ one finds
\begin{equation}\label{eqChargeJAdSCond1}
j^2<(s^2 +s +q^2)(3s^2 +s -q^2).
\end{equation}
On the other hand, the condition $\mu>\mu_*$ becomes
\begin{equation}\label{extrCond}
   \frac{9 \left(m^2- j^2\right)^2}{ \sqrt{6(\sqrt{X}- j^2-m^2)} \left(\sqrt{X}+2(j^2+ m^2)\right)}>1,
\end{equation}
where
\begin{equation}
    X=j^4+12 q^2 \left(m^2- j^2\right)^2+ 14 j^2 m^2+m^4 \quad \text{and} \quad X>0.
\end{equation}
Inserting $m$ from (\ref{eqm}) the condition (\ref{extrCond}) reduces to (\ref{eqChargeJAdSCond1}). Therefore, we consider \eqref{eqChargeJAdSCond1} as the sufficient condition for the existence of the KNAdS black hole. It is fulfilled above the blue surface depicted on all of the figures below.

\subsection{Global thermodynamic stability of Kerr-Newman-AdS black hole}

After identifying the region of existence we can study the conditions for global thermodynamic stability of the KNAdS black hole. The Hessian of the mass in $(s,j,q)$ space is defined by
\begin{equation}
\mathcal{H}=\left(\! \begin{array}{ccc}
\mathcal{H}_{ss} & \mathcal{H}_{sj} & \mathcal{H}_{sq} \\[5pt]
\mathcal{H}_{js} & \mathcal{H}_{jj} & \mathcal{H}_{jq} \\[5pt]
\mathcal{H}_{qs} & \mathcal{H}_{qj} & \mathcal{H}_{qq} 
\end{array} \!\right)=
\left(\!
\begin{array}{ccc}
\frac{\partial^2 m}{\partial s^2} & \frac{\partial^2 m}{\partial s \partial j} & \frac{\partial^2 m}{\partial s \partial q} \\[5pt]
\frac{\partial^2 m}{\partial j\partial s} & \frac{\partial^2 m}{\partial j^2} & \frac{\partial^2 m}{\partial j \partial q } \\[5pt]
\frac{\partial^2 m}{\partial q\partial s} & \frac{\partial^2 m}{\partial q \partial j} & \frac{\partial^2 m}{\partial q^2}
\end{array}
\!\right),
\end{equation}
The conditions for global thermodynamic stability require $\mathcal{H}$ to be positive definite matrix, i.e.
\begin{align}
&\mathcal{H}_{ss}>0, \quad \mathcal{H}_{jj}>0, \quad \mathcal{H}_{qq}>0, \quad \Delta_s>0, \quad \Delta_j>0, \quad \Delta_q>0, \quad \det\mathcal{H}>0,
\end{align}
where the explicit components of $\mathcal{H}$ and its principle minors $\Delta_{s,j,q}$ are listed in Appendix \ref{appA}. To analyze these conditions we will define the following functions:
\begin{align}\label{eqTBC}
&B(s,j,q)=A_1 A_2-j^2,
\\
\label{eqPCS}
&P(s,j,q)=A_1^3(3A_1-4s) +2j^2B_1 +j^4(3+4s),
\\\label{eqCOS}
&O(s,j,q)=A_1^4 \big[A_1 +2s(s\!-\!1)\big]  +2j^2A_1 B_2  +j^4 (A_1 +2q^2)(3+4s),
\\\label{RedSurface}
&R(s,j,q)=A_1^3 (3A_1-4s)(1+s) +2j^2A_1 \big[ q^2(1+3s) -s(1+s)^2\big]  -j^4(1+s),
\\\label{eqTGS}
&G(s,j,q)=A_1^4 \big[A_1 +2s(s\!-\!1)\big] (1\!+\!s) \! -2j^2A_1^2 \big[ A_1+sA_2-2s^3 \big]  \!-\!j^4 (A_1 +2q^2)(1\!+\!s) .
\end{align}
We choose their letters to signify the color of the surface they represent when equal to zero as depicted on the figures below, namely: $B$ -- blue, $P$ -- purple, $O$ -- orange, $R$ -- red, $G$ -- green.
Here we also define the following positive quantities:
\begin{align}
&A_1=s^2+s+q^2>0, 
\\
&A_2=3s^2+s-q^2>0, \\
&B_1=q^4 (3\! +\!2s) +2q^2 s(2\! +\!3s) + 3s^2 (1\!+\!s)^2 (1\!+\!2s)>0 , \\
&B_2= q^4(5+2s) +2q^2s(4s (s+2)+3) +3s^2 (1+s)^2(1+2s)>0 .
\end{align}

Let us first look at $H_{ss}$ given by \eqref{eqHss}. It is positive outside the purple closed surface, where $P>0$. On the other hand, $\mathcal H_{jj},\, \mathcal H_{qq}$ and $\Delta_s$ are positive everywhere, while $\Delta_j$ is positive outside the closed orange surface, where $O>0$, and $\Delta_q$ is positive above the red surface, where $R>0$. Finally, 
$\det\mathcal{H}$ is positive above  the green surface, where $G>0$. The region of global stability should be consistent with the region of existence (\ref{eqChargeJAdSCond1}) of the black hole (located above the blue curve $B>0$), which defines the region above the blue and the green surfaces:
\begin{align}\label{eqGlobalTDS}
&G(s,j,q)>0 \quad\text{and}\quad B(s,j,q)>0,
\end{align}
Therefore, one can achieve global equilibrium in the region above both blue and green surfaces. Below the green surface $(G<0)$  KNAdS can be considered only locally stable or unstable against fluctuations of the parameters.

\subsection{Local thermodynamic stability of  Kerr-Newman-AdS black hole}

The KNAdS black hole possesses various nontrivial heat capacities, and we will provide complete expressions for each of them below to illustrate their definitions.

\begin{figure}[H]
\centering
\includegraphics[scale=0.35]{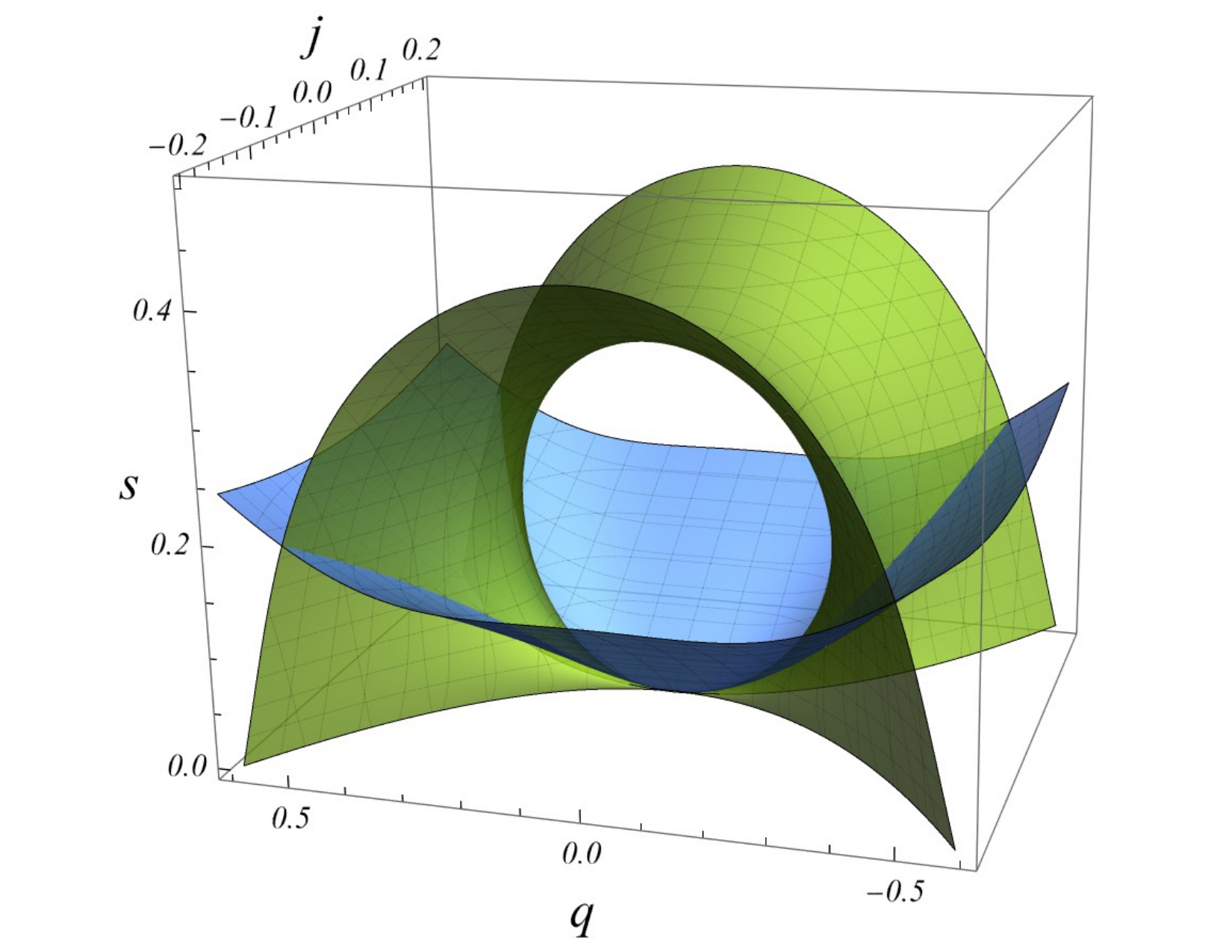}
\caption{ The blue surface represents the classically forbidden extremal case. Above the green surface the KNAdS black hole can be globally stable. One note also that the green surface corresponds to the Davies surface of $C_{\omega,\phi}$. }\label{Cm}
\end{figure}

First we consider all heat capacities at fixed mass $m$ in $(s,j,q)$ space:
\begin{align}
\label{CmOmega}
C_{m,\omega}&=\tau \frac{\partial s}{\partial \tau} \bigg|_{m,\omega} = \tau \frac{ \{s,m,\omega \}_{s,j,q} }{ \{ \tau,m,\omega \}_{s,j,q} } = \frac{ A_1 \big( A_1A_2-j^2 \big) s(1+s) }{ A_1^2(q^2+3s^2)(1\!+\!s) +j^2 \big[ sA_1 +q^2(1\!+\!2s) \big] } ,\\
\label{Cmj}
C_{m,j}&=\tau \frac{\partial s}{\partial \tau} \bigg|_{m,j} = \tau \frac{ \{s,m,j \}_{s,j,q} }{ \{ \tau,m,j \}_{s,j,q} } = \frac{A_1 \big( A_1 A_2 -j^2\big) s }{A_1^2 (q^2+3s^2) +j^2 \big[A_1 +s(1\!+\!2s) \big]},  \\
\label{CmPhi}
C_{m,\phi}&=\tau \frac{\partial s}{\partial \tau} \bigg|_{m,\phi} = \tau \frac{ \{s,m,\phi \}_{s,j,q} }{ \{ \tau,m,\phi \}_{s,j,q}} = \frac{ (A_1 +2q^2) \big(A_1A_2 -j^2\big)s(1+s) }{A_1 \big( B_2-2A_1^2 \big)  +j^2 (A_1+2q^2) (1\!+\!2s)}, \\
\label{Cmq}
C_{m,q}&=\tau \frac{\partial s}{\partial \tau} \bigg|_{m,q} = \tau \frac{ \{s,m,q \}_{s,j,q} }{ \{ \tau,m,q \}_{s,j,q} } = \frac{ \big(A_1A_2 -j^2\big) s(1+s)}{ q^4(1\!+\!2s) +2q^2s^2 +s^2(1\!+\!s)^2 (1\!+\!6s) +j^2(1\!+\!2s)}.
\end{align}
All of them are positive in the range of existence of the black hole \eqref{eqChargeJAdSCond1}, which is above the blue surface (Fig. \ref{Cm}). The local stability is defined in the region between the blue and the green surface ($B>0$ and $G<0$). Above these two surfaces ($B>0$ and $G>0$) the black hole can be globally stable (\ref{eqGlobalTDS}).

The next set of local heat capacities are defined at constant angular velocity together with fixed $j,\phi$ and $q$ correspondingly:
\begin{align}
C_{\omega,j}&=\tau \frac{\partial s}{\partial \tau} \bigg|_{\omega,j} \!\!=\!\tau \frac{ \{s,\omega,j \}_{s,j,q} }{ \{ \tau,\omega,j \}_{s,j,q} } \!=\! \frac{ A_1 \big( A_1A_2 -j^2\big) s(1+s)}{A_1^2 \!\big[ q^2(2\!+\!s) \!+\!s(1\!+\!s)(1\!+\!3s) \big] \!+\!j^2 \big[ q^2(2\!+\!3s) \!+\!s (1\!+\!s)^2 \big]}, \\
C_{\omega,\phi}&=\tau \frac{\partial s}{\partial \tau} \bigg|_{\omega,\phi} = \tau \frac{ \{s,\omega,\phi \}_{s,j,q} }{ \{ \tau,\omega,\phi \}_{s,j,q} } =\frac{2 A_1^3 \big(A_1A_2 -j^2\big) s(1+s)}{G(s,j,q)}, \\
C_{\omega,q}&=\tau \frac{\partial s}{\partial \tau} \bigg|_{\omega,q} = \tau \frac{ \{s,\omega,q \}_{s,j,q} }{ \{ \tau,\omega,q \}_{s,j,q}} =\frac{2A_1^2 \big(A_1A_2-j^2\big) s(1+s)}{R(s,j,q)}.
\end{align}
The heat capacity $C_{\omega,j}$ is positive in the range of existence (\ref{eqChargeJAdSCond1}). It has the same regions of local and global stability as in the case $m=const$, i.e. the local stability is between the blue and the green surface. The global equilibrium \eqref{eqGlobalTDS} can be achieved above these two surfaces (Fig. \ref{Cm}). 

The region of  positive $C_{\omega,\phi}$ coincides with the region of global stability (\ref{eqGlobalTDS}). Between the blue and the green surface ($B>0$ and $G<0$) the black hole is unstable ($C_{\omega,\phi}<0$). Above these two surfaces the black hole can be globally or locally stable.

\begin{figure}[H]
\centering
\includegraphics[scale=0.35]{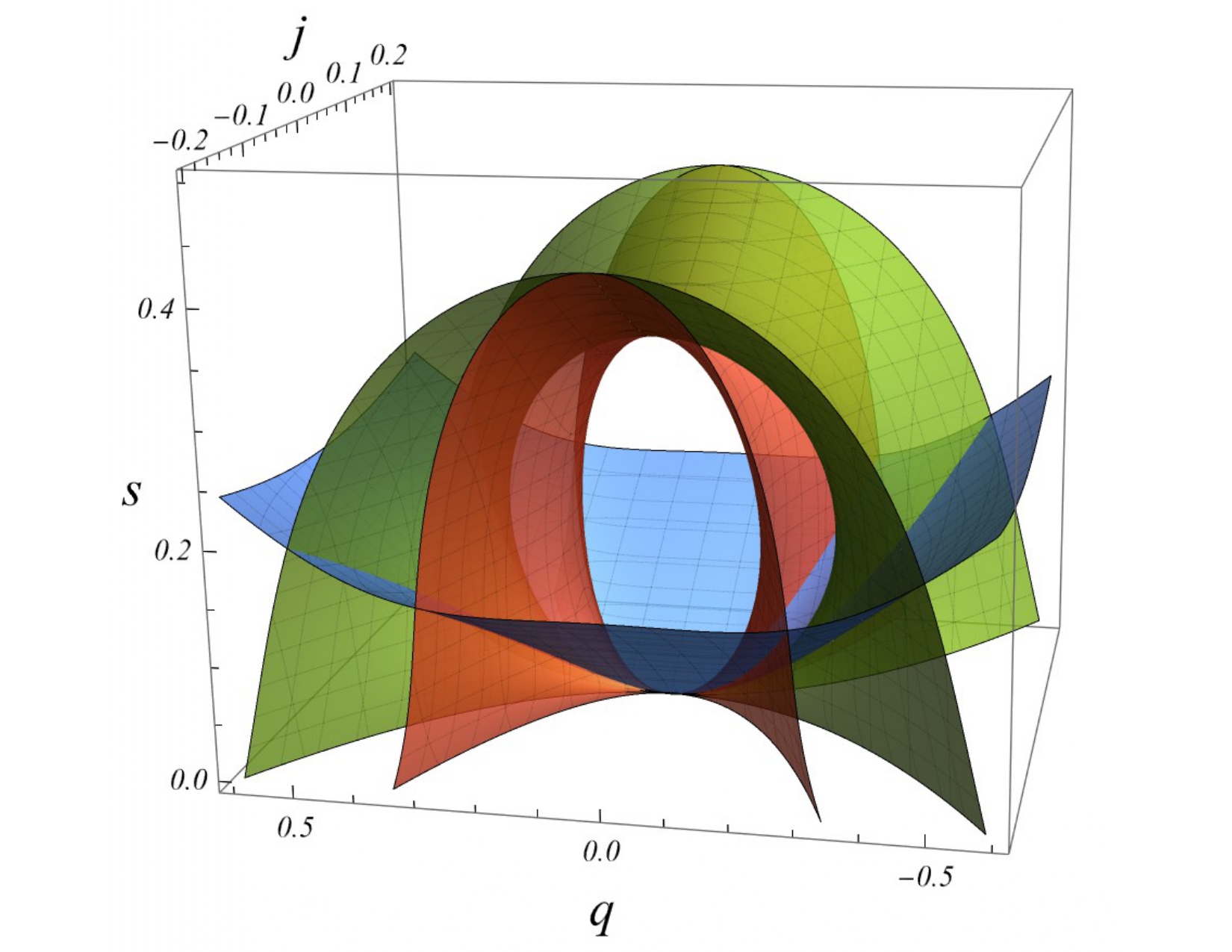}
\caption{Combined surfaces from Fig. \ref{Cm} and the Davies surface  for $C_{\omega,q}$ (the red surface). }\label{Cwq}
\end{figure}

Finally, the region of positive $C_{\omega,q}$ is above the blue and the red surface ($B>0$ and $R>0$). Between these two surfaces ($B>0$ and $R<0$) the black hole is unstable ($C_{\omega,q}<0$). In the region between the blue, the red and the green surfaces ($B>0, R>0$ and $G<0$), one has only local stability. Above the blue and the green surface \eqref{eqGlobalTDS} the black hole can be globally stable (Fig. \ref{Cwq}).

\begin{figure}[H]
\centering
\includegraphics[scale=0.35]{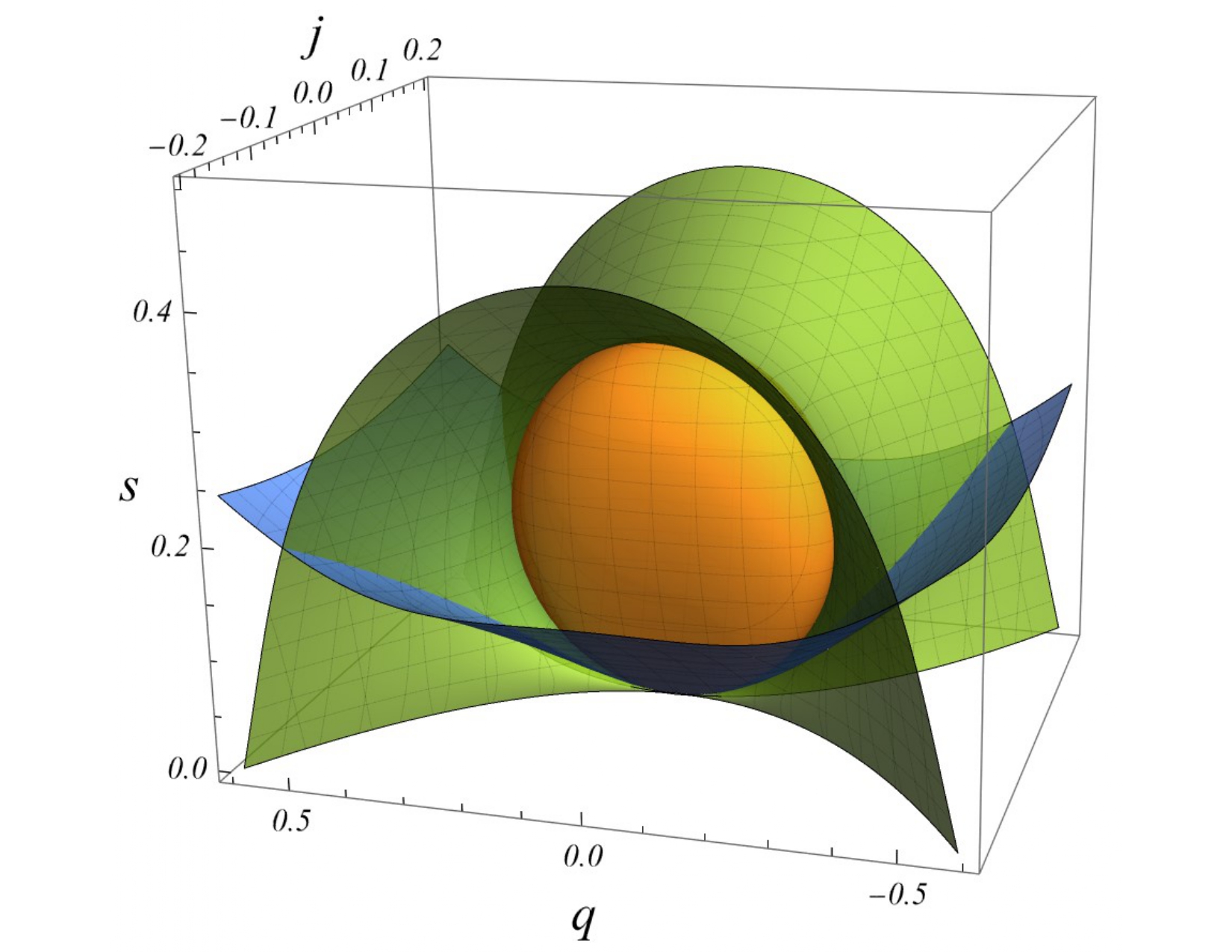}
\caption{Combined surfaces from Fig. \ref{Cm} and the Davies surface  for $C_{j,\phi}$ (the orange surface).}\label{Cjf}
\end{figure}

The next couple of heat capacities are given by
\begin{align}
C_{j,\phi}&=\tau \frac{\partial s}{\partial \tau} \bigg|_{j,\phi} = \tau \frac{ \{s,j,\phi \}_{s,j,q} }{ \{ \tau,j,\phi \}_{s,j,q}} =\frac{2s \big[ A_1^3 +j^2 (A_1+2q^2)(1\!+\!s) \big] \big(A_1A_2 -j^2\big)} {O(s,j,q)}, \\
C_{j,q}&=\tau \frac{\partial s}{\partial \tau} \bigg|_{j,q} = \tau \frac{ \{s,j,q \}_{s,j,q} }{ \{ \tau,j,q \}_{s,j,q} } = \frac{2s \big[ A_1^2 +j^2 (1+s)\big] \big(A_1A_2-j^2\big)} {P(s,j,q)}. 
\end{align}
One notes that $C_{j,\phi}>0$ is positive above the blue surface ($B>0$) and outside the closed orange surface ($O>0$). Inside the closed orange surface ($O<0$) the black hole is unstable ($C_{j,\phi}<0$). Between the blue and the green surface and outside the closed orange surface ($B>0, G<0$ and $O>0$) one has only local stability. Above the blue and the green surface ($B>0$ and $G>0$) the black hole can be in global equilibrium (Fig. \ref{Cjf}). 

\begin{figure}[H]
\centering
\includegraphics[scale=0.35]{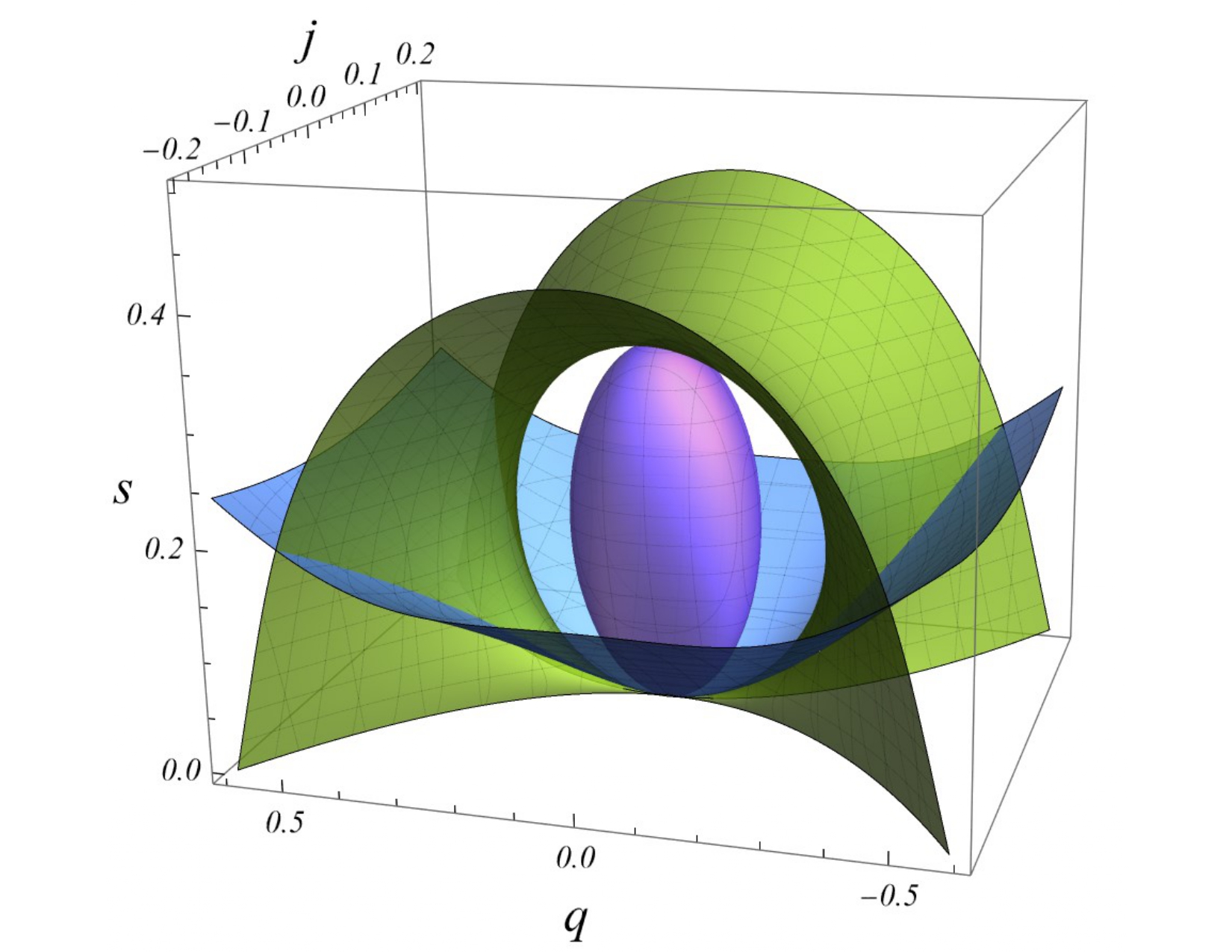}
\caption{Combined surfaces from Fig. \ref{Cm} and the Davies surface for $C_{j,q}$ (the purple surface). }\label{Cjq}
\end{figure}

The heat capacity $C_{j,q}$ is positive above the blue surface and outside the closed purple surface ($B>0$ and $P>0$). Inside the closed purple surface ($P<0$) the black hole is unstable  ($C_{j,q}<0$). Between the blue and the green surface and outside the closed purple surface ($B>0, G<0$ and $P>0$) one has only local stability. Above the blue and the green surface \eqref{eqGlobalTDS} the black hole can be in globally stable (Fig. \ref{Cjq}),

The final admissible heat capacity is given by $C_{\phi,q}$:
\begin{align}
C_{\phi,q}&\!=\!\tau \frac{\partial s}{\partial \tau} \bigg|_{\phi,q}\!\! =\! \tau \frac{ \{s,\phi,q \}_{s,j,q} }{ \{ \tau,\phi,q \}_{s,j,q}} = \frac{A_1 \big( A_1A_2 -j^2\big) s(1+s)}{A_1^2 \big[ q^2 (2\!+\!s)\!+\!s (1\!+\!s)(1\!+\!3s) \big]\! +\!j^2 \big[ q^2(2\!+\!3s) \!+\!s(1\!+\!s)^2 \big]},
\end{align} 
which coincides with $C_{\omega,j}$.

In summary, all of the heat capacities are positive in the region of global stability: above the blue and the green surfaces ($B>0$ and $G>0$). However, we showed that below the green surface there are several regions of local thermodynamic stability, defined by the properties of the corresponding heat capacities. This concludes the full classical local and global thermodynamic stability analysis of the KNAdS black hole.

\section{Conclusion}

Our investigation thoroughly explored  the thermodynamic stability of AdS black holes, employing the robust Sylvester criterion within the mass-energy representation. This criterion, grounded in the strict convexity property of the energy potential in thermodynamics, has illuminated the boundaries of classical thermodynamics when applied to AdS black hole systems. We showed that AdS black holes exhibit regions of global stability, primarily influenced by the presence of the cosmological parameter $l$, effectively stabilizing the system. Moreover, by setting $l \to \infty$, we can recover the results of \cite{Avramov:2023eif}.

Utilizing the Nambu bracket formalism for local heat capacities, we have identified regions of local stability for thermodynamic processes, corresponding to specific restrictions imposed on the system. In the AdS-Schwarzschild scenario, the region of local thermodynamic stability aligns with the global one, albeit achievable only for black holes with event horizons comparable to the AdS radius. However, in other black hole cases such as RNAdS, Kerr-AdS, and KNAdS, regions of local stability extend beyond the boundaries of the global one.

While some of our findings may have already been presented in existing literature, they have been scattered across various texts\footnote{See for example \cite{Dolan:2014lea, Dolan:2013yca, Dolan:2010ha, Dolan:2011jm, Dolan:2014jva, Carlip:2003ne, Carter:2005uw, Sinha:2015zxv, Sinha19aaa, Sinha21bb,
Sinha:2020uwf, Sinha:2016evq, Sinha:2017obh, Sahay:2010wi, Banerjee:2010bx, Liu:2010sz, Tsai:2011gv, Roychowdhury:2013klt, Kubiznak:2014zwa,Peca:1998cs}.}. Nonetheless, to our knowledge, a comprehensive classical analysis of the thermodynamic stability of AdS black hole systems has not been presented with such detail before. Our approach underscores the applicability of classical thermodynamics to black hole physics, while acknowledging the potential for deviations from classical setups that could influence the expectations of classical observables. Despite this, we anticipate classical thermodynamics to remain valid even in many exotic or complex cases due to its inherent rigidity.

About further developments, we are advancing  to explore the impact of
converting the cosmological constant into a thermodynamic parameter and it
will be the focal point of the forthcoming third installment of our
research. Additionally, investigating generic stability criteria for other black hole solutions, including hairy black holes, lower and higher-dimensional black holes, regular black holes, black holes with extended thermodynamics, and related systems, holds significant promise for future inquiry.

\section*{Acknowledgments}

The authors would like to express their gratitude to S. Yazadjiev for his invaluable comments and discussions. H. D. thankfully acknowledges the support by the program “JINR-Bulgaria” of the Bulgarian Nuclear Regulatory Agency. V. A. gratefully acknowledges the support by the Simons Foundation and the International Center for Mathematical Sciences in Sofia for the various annual scientific events. M. R., R. R. and T. V. were fully financed  by the European Union- NextGeneration EU, through the National Recovery and Resilience Plan of the Republic of Bulgaria, project BG-RRP-2.004-0008-C01.

 \appendix

 \section{Hessian of the mass of the KNAdS black hole}\label{appA}

The Hessian of the mass in $(s,j,q)$ space is
\begin{equation}
\mathcal{H}=\left(\! \begin{array}{ccc}
\mathcal{H}_{ss} & \mathcal{H}_{sj} & \mathcal{H}_{sq} \\[5pt]
\mathcal{H}_{js} & \mathcal{H}_{jj} & \mathcal{H}_{jq} \\[5pt]
\mathcal{H}_{qs} & \mathcal{H}_{qj} & \mathcal{H}_{qq} 
\end{array} \!\right)=
\left(\!
\begin{array}{ccc}
\frac{\partial^2 m}{\partial s^2} & \frac{\partial^2 m}{\partial s \partial j} & \frac{\partial^2 m}{\partial s \partial q} \\[5pt]
\frac{\partial^2 m}{\partial j\partial s} & \frac{\partial^2 m}{\partial j^2} & \frac{\partial^2 m}{\partial j \partial q } \\[5pt]
\frac{\partial^2 m}{\partial q\partial s} & \frac{\partial^2 m}{\partial q \partial j} & \frac{\partial^2 m}{\partial q^2}
\end{array}
\!\right).
\end{equation}
with explicit components written by:
\begin{align}\label{eqHss}
&\mathcal{H}_{ss}= \frac{ A_1^3 (3A_1-4s) +2j^2 B_1  +j^4(3+4s) } {4 \sqrt{ s^5 \big[  A_1^2 +j^2(1+s) \big]^3\, }}, \\
&\mathcal{H}_{jj}=\frac{ A_1^2(1+s) } {\sqrt{s\, \big[ A_1^2 +j^2(1+s) \big]^3\,}},  \\
&\mathcal{H}_{qq}= \frac{2\big[ A_1^3 +j^2(A_1 +2q^2)(1+s) \big]} {\sqrt{s\, \big[ A_1^2 +j^2(1+s) \big]^3\, }}, \\
&\mathcal{H}_{sj}= -\, \frac{ j A_1 \big[ A_1 +sA_2 +2s(1+2s) \big]  +j^3 (1+s)} {2\sqrt{ s^3 \big[ A_1^2 +j^2(1+s) \big]^3\, }}, \\
&\mathcal{H}_{sq}= -\, \frac{q \big[ A_1^3 -j^2(A_2-2s^2)(1+2s) \big]} { \sqrt{s^3 \big[ A_1^2 +j^2(1+s) \big]^3\,}}, \\
&\mathcal{H}_{jq}= -\,\frac{2jqA_1(1+s) }{\sqrt{ s\big[ A_1^2 +j^2(1+s) \big]^3 }}.
\end{align}
For convenience we have defined the following notations:
\begin{align}
&A_1=s^2+s+q^2, \quad  A_2=3s^2+s-q^2, \\
&B_1=q^4 (3\! +\!2s) +2q^2 s(2\! +\!3s) + 3s^2 (1\!+\!s)^2 (1\!+\!2s) , \\
&B_2= q^4(5+2s) +2q^2s(4s (s+2)+3) +3s^2 (1+s)^2(1+2s) .
\end{align}
The principle minors are defined by
\begin{equation}
\Delta_s=\left|\! \begin{array}{cc}
\mathcal{H}_{jj} & \mathcal{H}_{jq} \\[5pt]
\mathcal{H}_{qj} & \mathcal{H}_{qq}  
\end{array} \!\right|, \quad  \Delta_j=\left|\! \begin{array}{cc}
\mathcal{H}_{ss} & \mathcal{H}_{sq} \\[5pt]
\mathcal{H}_{qs} & \mathcal{H}_{qq}  
\end{array} \!\right|, \quad  \Delta_q=\left|\! \begin{array}{cc}
\mathcal{H}_{ss} & \mathcal{H}_{sj} \\[5pt]
\mathcal{H}_{js} & \mathcal{H}_{jj}  
\end{array} \!\right|.
\end{equation}
where one explicitly has 
\begin{align}
&\Delta_s=\frac{ 2A_1^3 (1+s)}{s \big[ A_1^2 +j^2(1+s) \big]^2}, \\
&\Delta_j= \frac{ A_1^4 \big[A_1 +2s(s-1)\big] +2j^2A_1B_2 +j^4(A_1+2q^2)(3+4s) }{2 s^3 \big[ A_1^2 +j^2(1+s) \big]^2}, \\
&\Delta_q= \frac{ A_1^3(3A_1-4s)(1+s)  +2j^2A_1 \big[ q^2(1+3s) -s(1+s)^2 \big] 
 -j^4(1+s) }{4s^3 \big[ A_1^2 +j^2 (1+s) \big]^2}.
\end{align}
Finally, the determinant of the Hessian evaluates to
\begin{align}
\det\mathcal{H}= \frac{   A_1^4 \big[A_1 +2s(s\!-\!1)\big] (1\!+\!s)  -2j^2 A_1^2 \big[ A_1 +sA_2 -2s^3 \big]  -j^4(A_1 +2q^2)(1\!+\!s)}  { 2\sqrt{s^7 \big[ A_1^2 +j^2(1+s) \big]^5\, }}.
\end{align}

\section{Section planes in the state space of KNAdS}\label{secB}

\subsection{Recovering RNAdS by $j=0$ section}

If we superimpose Fig. \ref{Cwq} and Fig. \ref{Cjq}, it becomes evident that the red surface and the purple closed surface come into contact along a single ellipse situated in the $j=0$ plane. This ellipse precisely corresponds to the orange ellipse (\ref{OrangeEllipcse}), illustrated in Figure \ref{RNAdS} in the RNAdS scenario, residing at the intersection of the $j=0$ plane traversing the red and purple surfaces.

Examining Fig. \ref{Cjf}, within the same $j=0$ intersection plane, we observe that the green and orange surfaces intersect at a single ellipse. This ellipse corresponds to the green ellipse (\ref{GreenEllipse}) featured in Figure \ref{RNAdS} in the RNAdS context.

Finally, the blue surface (visible in all figures) intersects the $j=0$ plane at the blue hyperbola depicted in Fig. \ref{RNAdS}.

\subsection{Recovering Kerr-AdS by $q=0$ section}

Examining the $q=0$ plane, one observes that the green and blue surfaces intersect this plane, forming the green and blue curves as illustrated in Fig. \ref{KerrAdS}.

Upon merging Fig. \ref{Cjf} and Fig. \ref{Cjq}, it becomes apparent that the orange closed surface and the purple closed surface come into contact, forming a closed orange curve depicted in Fig. \ref{KerrAdS}.

\bibliographystyle{utphys}
\bibliography{References}
\end{document}